\documentclass[preprint,12pt]{elsarticle}

\usepackage{amsmath,amssymb,amsfonts}
\usepackage{manyfoot}%
\usepackage{booktabs}%
\usepackage{algorithm}%
\usepackage{algorithmicx}%
\usepackage{algpseudocode}%
\usepackage{listings}%
\usepackage{amsthm}%
\usepackage{mathrsfs}%
\usepackage{xcolor}%
\usepackage{textcomp}%
\usepackage{graphicx}%
\usepackage{hyperref}
\usepackage{subcaption}
\hypersetup{
	colorlinks=true,
	linkcolor=blue,
	filecolor=magenta,      
	urlcolor=green,
}
\makeatletter
\def\ps@pprintTitle{%
	\let\@oddhead\@empty
	\let\@evenhead\@empty
	\let\@oddfoot\@empty
	\let\@evenfoot\@oddfoot
}
\makeatother

\begin{document}

\begin{frontmatter}



\title{Forecasting dengue outbreaks with uncertainty using seasonal weather patterns }


\author[inst1]{Piyumi Chathurangika}

\affiliation[inst1]{organization={Research \& Development Center for Mathematical Modeling, Department of Mathematics},%Department and Organization
            addressline={Faculty of Science}, 
            city={University of Colombo},
            %postcode={00000}, 
            %state={State One},
            country={Sri Lanka}}

\author[inst1]{S.S.N. Perera}
\author[inst1]{Kushani De Silva \corref{cor1}}
\cortext[cor1]{Corresponding author}

%\affiliation[inst2]{organization={Department Two},%Department and Organization
%            addressline={Address Two}, 
%            city={City Two},
%            postcode={22222}, 
%            state={State Two},
%            country={Country Two}}

\begin{abstract}
Dengue is a vector-borne disease transmitted to humans by vectors of genus Aedes and is a global threat with health, social, and economic impact in many of the tropical countries including Sri Lanka. The virus transmission is significantly impacted by environmental conditions, with a notable contribution from elevated per-capita vector density. These conditions are dynamic in nature and specially having the tropical climate, Sri Lanka experiences seasonal weather patterns dominated by monsoons. In this work, we investigate the dynamic influence of environmental conditions on dengue emergence in Colombo district where dengue is extremely prevalent in Sri Lanka. A novel approach leveraging the Markov chain Monte Carlo simulations has been employed to identify seasonal patterns of dengue disease emergence, utilizing the dynamics of weather patterns governing in the region. The newly developed algorithm allows us to estimate the timing of dengue outbreaks with uncertainty, enabling accurate forecasts of upcoming disease emergence patterns for better preparedness.
\end{abstract}




\end{frontmatter}


\section{Introduction}
\label{sec:sample1}
	Dengue fever is an infectious disease caused by the vector-borne dengue virus (DENV-1 to DENV-4) transmitted by the Aedes mosquitoes particularly Aedes aegypti \citep*{halstead1988pathogenesis}. The fever caused by at least one of the aforementioned serotypes is a major global health concern with an estimated 100 million cases occurring worldwide every year. The cases of dengue fever have been increased in recent years, 30-fold over the past 50 years with a 50\% increase in the past decade alone, with a growing number of countries reporting outbreaks \citep*{bhatt2013global, guzman2010dengue, CDTR2023}. In 2020 alone, over 2.5 million cases of dengue were reported worldwide with 1,018 (0.04\%) deaths \citep{who_2021}. The World Health Organization (WHO) has declared dengue fever an endemic in over 100 countries, mostly in tropical and subtropical regions of the world, including Southeast Asia, South America, and parts of Africa. The dengue threat is highest in Asia and the Americas, with Southeast Asia and the Western Pacific region accounting for over $70\%$ of the global dengue disease statistics.

Dengue is one of the most stressing health concerns in Sri Lanka, particularly reporting high numbers in the western province with outbreaks occurring frequently \citep{mohlk2019}. In recent years Sri Lanka has seen an increase in the dengue cases with an unexpected high number of cases being reported in 2017 and 2019. According to the Epidemiology Unit of Sri Lanka, in 2019 there were over 51,000 reported cases, with over 90 deaths (0.17\%) whereas in 2017 there were 186,101 reported cases with over 440 (0.23\%) deaths collapsing the Sri Lankan health system \citep{epidemiologyunit}. The Sri Lankan government has implemented various health and social measures to prevent and control the spread of the disease by conducting public awareness campaigns, clean-up programs to remove mosquito breeding sites, insecticide fumigation and increasing the use of mosquito repellents \citep{mohlk2019, national2019}. However, as in many other tropical countries, Sri Lanka too has failed to completely wipe-out or successfully control the disease emergence. Hence, identifying regional-dependent factors affecting the disease emergence and transmission is of utmost importance to curtail this serious health issue and strengthen the health system. 

The human plays the role of the host to this deadly virus, while both the Aedes vector and the human act as carriers to the virus. Thus, the population densities of both human and vector are intricately linked to the virus transmission \citep{padmanabha2012interactive}. The Aedes vectors thrive in warm and humid environments commonly found in tropical and subtropical regions \citep{liu2014vectorial}. In particular, the Aedes aegypti has multiple stages in their life cycle - egg, larva, pupa, and adult. Mosquitoes lay their eggs in stagnant clean water sources such as uncovered containers, discarded tires, and water-filled areas. With an abundant supply of breeding sites and favorable climatic conditions, the vector population can grow rapidly. Consequently, this high vector population density can amplify the likelihood of dengue transmission for several reasons; (1) a larger vector population means a greater chance of female vectors being infected with the dengue virus, (2) more mosquitoes increase the probability of contact between infected vectors and susceptible individuals enabling the virus to be transmitted more rapidly  \citep{focks2003review}.  Therefore, the vector population has a significantly high impact on dengue transmission within human population. 
%Additionally, since the climatic factors impact the growth of the mosquito population, it is a major contributing factor to the spread of dengue in Sri Lanka, as rising temperatures and cyclic rainfall patterns can create breeding sites and favorable conditions for mosquito breeding \citep{moh_lk_2019}. 

The transmission of dengue is closely linked to environmental conditions as it has a direct impact on dengue mosquito population density \citep{mohlk2019,naish2014climate,wmp_2019}. The warmer and humid environments,  typically between $25-30^0$C ($77-86^0$F) of temperatures and above 60\% of humidity levels, provide a highly favorable environment for the Aedes mosquitoes and thus the virus emergence happen very rapidly \citep{tjaden2013extrinsic}. This favorability of climatic conditions for dengue disease emergence is discussed in many studies (E.g. \citep{wang2019combination, jourdain2020importation, robert2019temperature}). Sri Lanka experiences an average temperature ranging from $26$ to $30^0C$, with coastal regions being slightly warmer than inland areas. The relative humidity remains consistently above 70\% throughout the year, reaching nearly 90\% during the monsoon season in June. These conditions create highly favorable breeding sites for Aedes mosquitoes \citep{wb_2020}. With the seasonal monsoons experienced in Sri Lanka, rainfall is one of the major factors that provide favourable conditions for breeding and transmission of the disease \citep{li1985rainfall, environments6060071, chathurangik2022dengue}. Although Sri Lanka has fairly steady weather conditions, the dynamic behaviour of rainfall play a major role in disease transmission. On top of these dynamics that causes rapid transmission, the host, their high density and their mobility play the role of elevating the transmission rate. Overall, these environmental favourability for densed mosquito population, densed human population and human mobility are all present in Colombo district and that resulted in high severity of the disease in Colombo.

%With the environmental conditions that are influencing virus breeding, rainfall plays a significant role in Sri Lanka due to its drastic dynamics compared to subtle  changes observed in temperature and humidity \cite{chathurangik2022dengue}. In a favorable environment where the mosquitoes have apple opportunity to spread the virus among humans, human behavioural changes play a role in elevating or lowering the disease emergence patterns. For instance, enhanced human mobility to areas with high disease emergence may increase the number of dengue infections further. On the other hand, dengue prevention programs can reduce the expected disease emergence counts. 

Due to the severity of dengue disease, its prediction is very important to handle the strain on the Sri Lankan health system. Moreover, rapid development of more sensitive early dengue detection, diagnostics and surveillance measures are vital for better handling of the dengue threat \citep{tambo2016outwitting}. Prediction of dengue emergence has been done using many methods in various studies using different viewpoints \citep{louis2014modeling, martheswaran2022prediction, leung2023systematic}. The weather conditions and dengue emergence has already been discussed in many studies conducted in Sri Lanka \citep{faruk2022impact, edirisinghe2017contribution, sun2017spatial, ehelepola2015study}. Further, these studies have identified rainfall and temperature as the environmental factors that has the most significant impact on dengue emergence. Moreover, the impact of these factors on dengue emergence is found to be seasonal, resulting the dynamic behavior of dengue emergence within a region. However, the relationship between these seasonal weather parameters and dengue emergence patterns have not yet been properly quantified. Hence analyzing the dynamical behaviour of both dengue and weather variables simultaneously is of utmost importance. In this study, rainfall has been identified as the governing weather variable on dengue emergence and its seasonal variation is taken into account to analyze the dynamical behaviour of dengue emerging patterns. The impact of seasonal rainfall is captured through the vector population from which the dengue infected human population is estimated. The relationship between the vector, host and rainfall are developed using a system of ordinary differential equations. Moreover, the uncertainty of the estimations of dengue has been quantified using Bayesian inferencing with Markov chain Monte Carlo simulations.

The next section of this paper explains background of this study including the study-area data collection and the study period. Section 3 of this paper elaborates the methodology followed together with the detailed analysis. The results obtained and the discussion based on the results are explained in the last section of the paper. 

	\section{Background of the study}
\begin{figure}[ht]
	\hfill
	%\subfigure[]{\includegraphics[width=0.3\textwidth]{population.jpg}}  \hfill
	%\subfigure[]{\includegraphics[width=0.65\textwidth]{dengue_by_districts.jpg}}
	\includegraphics[width=\textwidth]{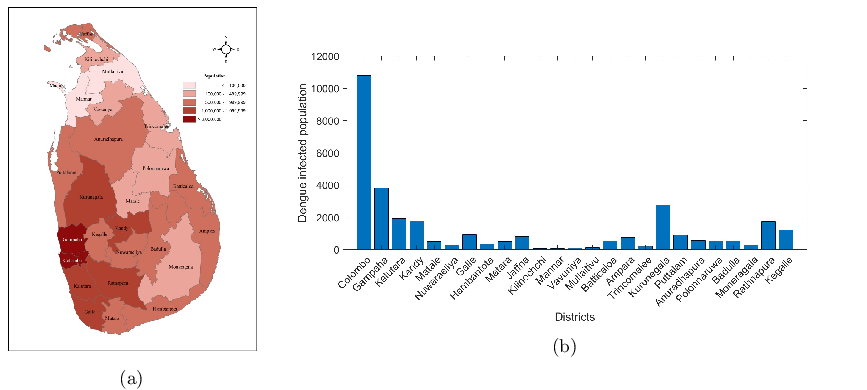} 
	\hfill
	\caption{(a) The district population density distribution and (b) infected dengue distribution for the year 2012 \citep{mohlk2019, statisticsSL}.}
	\label{fig_1}
\end{figure}

Sri Lanka has highly densed population in the Colombo district and as a result it continuously accounts for high number of reported dengue cases (see Fig. \ref{fig_1}b). The Colombo Municipal Council (CMC) area within the district exhibits the highest population density, with 24,857 residents per $km^2$, making it the most densely populated region. Within the district, Colombo Municipal Council (CMC) area has a population of 24,857 per $km^2$ and is the highest densed area \citep{colombo_website,comtrans}.  %\href{https://power.larc.nasa.gov/data-access-viewer/}{NASA Webpage}.

The rainfall pattern in Sri Lanka is heavily influenced by the two monsoons that affect the island. These are the southwest monsoon, lasts from May to September, and the northeast monsoon, lasts from December to February \citep{tjaden2013extrinsic,wb_2020,karunathilaka2017changes}. 
During the southwest monsoon, the southwestern part of the country, including Colombo and the central highlands receive the heaviest rainfall. The average total rainfall during this period is between 2500-4000 mm. During the northeast monsoon, the eastern and northeastern parts of the country receive the heaviest rainfall. The average rainfall during this period is between 1000-2000 mm. The southwestern part of the country experiences relatively dry weather during this time, reporting the lowest rainfall of 170 mm in Chilaw.

\begin{figure}[ht]
	\hfill
	%\subfigure[]{\includegraphics[width=0.48\textwidth]{dengue_variation.eps}}
	%\hfill
	%\subfigure[]{\includegraphics[width=0.48\textwidth]{rainfall_variation.eps}}
	%\hfill
	\includegraphics[width=\textwidth]{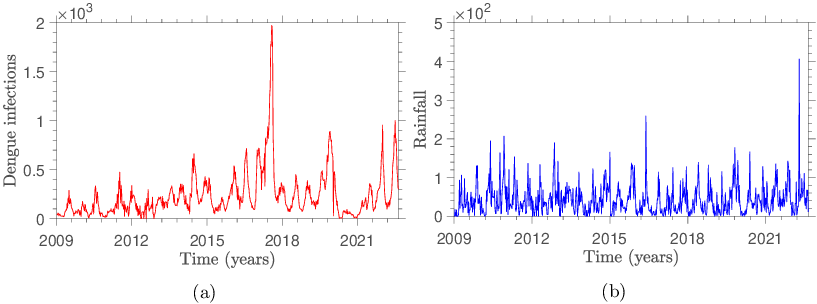}
	\caption{Variation of (a) dengue infected human population and (b) weekly rainfall in the CMC area from 2009-2022.}
	\label{fig_2}
\end{figure}

Over the course of a year, observable cyclic patterns of dengue emergence can be noted. These cycles align with the monsoon seasons mentioned earlier, as illustrated in Fig. \ref{fig_2}. Depending on cyclic pattern of dengue emergence, a single year can be characterized into four seasons where two of them shows increasing dengue emergence patterns and the other two seasons show decreasing dengue emergence patterns. 

Moreover, the impact of rainfall on emergence of dengue is studied through wavelet analysis in \citep{chathurangik2022dengue,SANTOS2019794,Wavelet} which show that there is a 10-week time lag between the occurrence of rainfall and emergence of dengue in the CMC area. {With the time lags identified, the total number of seasons identified were 54 that starts with second season of 2009 and ends with third season of 2022. The time window of each of these 54 seasons (4 identified seasons per year) were identified based on the increasing and decreasing trends in the reported rainfall fluctuations. Consequently, the number of weeks comprising each season may differ, resulting in a distinctive duration for each one.} Thus, the time span of a considered season $i$ is denoted by $t(i)$.

	\section{Method and Analysis}
\begin{figure}[ht]
	\centering
	\includegraphics[width=6cm]{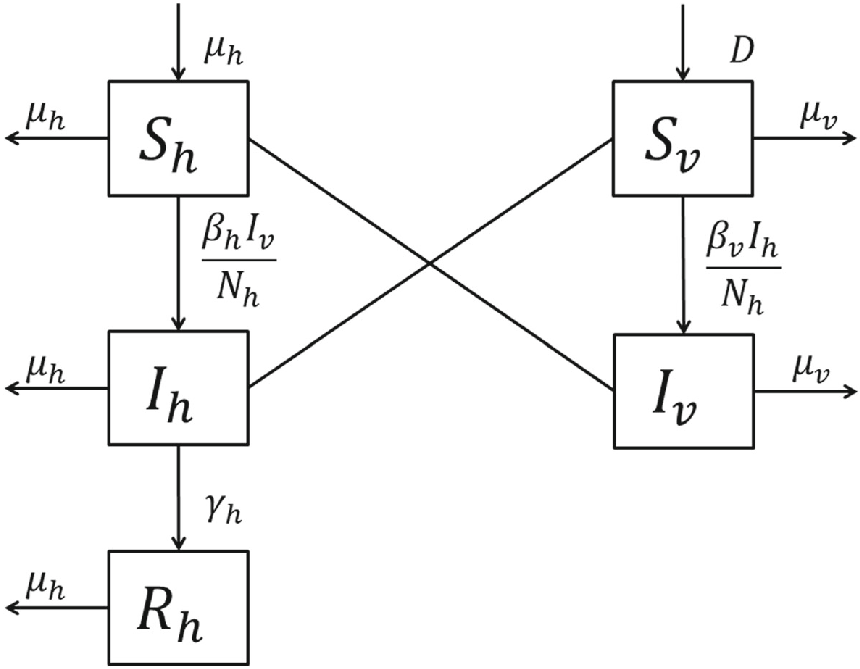}
	\caption{Schematic diagram: transmission of the dengue virus between human and vector populations. $S_h$, $I_h$, and $R_h$ are the susceptible, infected and recovered human populations respectively. $S_v$ and $I_v$ are the susceptible and infected vector (mosquito) populations respectively. }
	\label{fig_3}
\end{figure}

In this section we discuss the methodology and data analysis used to estimate and forecast the dengue disease emergence in the CMC area. In Fig. \ref{fig_3}, the dengue transmission schematic diagram used in this study is depicted, in which five compartments represent the human and vector populations respectively \citep{Erandi2021}. 

Taking the total populations, i.e. $N_h=S_h+I_h+R_h$ and $N_v=S_v+I_v$, the five compartmental model (representing five state variables in ODE system) can be reduced to three state variables. The resulting ODE system of $I,R,$ and $V$ are given in model \eqref{3D_syst}. The definitions of the parameters of model \eqref{3D_syst} are given in Table \ref{tab1}. 
\begin{align}
	\label{3D_syst}
	\begin{cases}
		\vspace*{2mm}
		\dfrac{dI}{dt}&=\beta_h z  V(1-I-R)-(\mu_h+\gamma_h)I,\\
		\vspace*{2mm}
		\dfrac{dR}{dt}&=\gamma_h I-\mu_h R, \\
		\vspace*{2mm}
		\dfrac{dV}{dt}&=\beta_v(1-V)I-\mu_v V.
	\end{cases}
\end{align}
Of the parameters given in Table \ref{tab1}, the per-capita vector density is directly dependent on rainfall patterns due to the abundant mosquito breeding sites emerging after rainfall \citep{sirisena2017effect, rainf, chathurangik2022dengue}. {Hence, the mosquito population is directly proportional to the rainfall patterns in the selected region of study. Thus the relationship in Eq.\eqref{z_with_Rf} is assumed since there is no evidence to add any other complexity to it. The per-capita vector density for a season can be given by,} 
\begin{align}
	\label{z_with_Rf}
	\mathbf{z}=a \mathbf{R_f} ,
\end{align}
\noindent {where $\mathbf{R_f}$ is the vector of reported weekly total rainfall per season, $a$ is the rainfall coefficient per season and $\mathbf{z}$ is the vector of per-capita vector density per season. From this point onward, the bold symbols such as $\mathbf{R_f}$ represent vectors, whereas non-bold symbols represent scalars. Moreover, the subscript $i$ shows the seasonal index of a given variable. According to the assumption that the per-capita vector density is directly proportional to seasonal rainfall, the corresponding proportionality constant is taken as our parameter $a$. It is a scalar defined for each season ($a_i$ for each $i$) and estimated in our study based on seasonal rainfall patterns. Since this parameter reflects how rainfall influences the dynamics of per-capita vector density for a given season, $a_i$ is interpreted as the seasonal rainfall coefficient.}
\begin{table}[ht]
	\centering
	\caption{Variables and parameter descriptions of model \eqref{3D_syst}. The bold-faced letters given are the vectors within a period of identified season. Other variables are scalars within a season.}
	\label{tab1}
	\begin{tabular}{@{}llll@{}}
		\toprule
		{Variable/Parameter} & {Description}             \\ 
		\midrule
		$\mathbf{I}$  & Infected human population density $I\in [0,1]$ \\   
		$\mathbf{R}$ & Recovered human population density $R\in [0,1]$  \\
		$\mathbf{V}$   & Infected vector population density $V\in [0,1]$ \\
		$\mathbf{z}$  & Per-capita vector density  \\
		$\beta_h$      & Transmission rate from vector to host       \\
		$\mu_h$       & Host death rate                             \\
		$\gamma_h$     & Host recovery rate                               \\
		$\beta_v$       & Transmission rate from host to vector     \\
		$\mu_v$             & Vector death rate\\
		$\mathbf{R_f}$  & Weekly total rainfall per season  \\
		$a$ & Seasonal rainfall coefficient	\\
		\bottomrule
	\end{tabular}
\end{table}

The simulation of model \eqref{3D_syst} requires the estimation of seasonal $\mathbf{z}$ based on rainfall which then requires the estimation of $a$. We employ the dynamic parameter estimation with Bayesian setup using Markov chain Monte Carlo. The Bayes' rule is given below,
\begin{subequations}\label{eq. bayes}
\begin{align} 
	p\left( \theta|x\right) &=\dfrac{p\left(\theta \right)p\left( x|\theta\right)  }{\int p\left(\theta \right)p\left( x|\theta\right) d\theta }, \label{3a}\\
	&=K p\left(\theta \right)p\left( x|\theta\right),  \label{3b}
\end{align}
\end{subequations}
where $\theta$ and $x$ represent the parameters and data respectively with $K$ representing the evidence that normalizes the posterior probability distribution. In this problem, parameters (unknowns to be estimated) and data (observed information) are identified in Table \ref{tab2}. One of the advantages of using Bayesian setup is its ability to infer the error variance $\sigma$. In our estimation process we are inferring the rainfall coefficient $a$ and error variance $\sigma$ for each season. Here, the error vector for each season is given by,
\begin{align}
	\mathbf{\epsilon} =\mathbf{\hat{I}} - \mathbf{I_{obs}},
\end{align}
where $\mathbf{\hat{I}} = \displaystyle\int \dfrac{dI}{dt}dt$ and $\mathbf{I_{obs}}$ is the vector of observed dengue incidences for a season. 
In the process of parameter estimation, the actual dengue data are compared against the solutions of the ODE system in Eq.\eqref{3D_syst}. During these simulations, the recorded rainfall data ($\mathbf{R_f}$) and dengue incidence data ($\mathbf{I_{obs}}$) are fed into the system. Weekly reported dengue cases in the CMC area are obtained from the National Dengue Control Unit of Sri Lanka. The rainfall data required for this study were extracted from \cite{NASAPOWER}. In Table \ref{tab2}, the parameters to be estimated are identified. Further, the remaining dengue-related parameters ($\mathbf{P}$) are borrowed from existing literature, as they are not directly measurable and are not the primary focus of this work.
\begin{table}[ht]
	\centering
	\caption{Observables and unknown parameters for the Bayesian paradigm. All the parameters given in the table are for a season. The definitions of the terms can be found in Table \ref{tab1}. Here $p$ is the number of seasons for the study period.}
	\begin{tabular}{@{}llll@{}}
		\toprule
		Description& Notation\\ 
		\midrule 
		Parameters & $a,\sigma$ \\
		Data & $\mathbf{I_{obs}},\mathbf{R_f}, p$\\
		Literature data & $\mathbf{P} = \left[ \beta_h,\mu_h,\gamma_h,\beta_v, \mu_v \right] $\\
		Initial values & $\mathbf{Y_0} =\left[I_0,R_0,V_0,a_0 \right] $\\
		\bottomrule
	\end{tabular}
	\label{tab2}
\end{table}

The Bayesian rule can be applied to the data and parameters of the model \eqref{3D_syst} for a season as below,
%\begin{align}
%	&p\left( a| \mathbf{I_{obs}},\mathbf{R_f}, \mathbf{P},\mathbf{Y_0},\mathbf{\sigma}, \mathcal{I}\right) \notag\\
%	&\propto p\left( a|\mathbf{R_f},\sigma, \mathcal{I}\right)p\left(\mathbf{R_f}|\sigma,\mathcal{I} \right) p\left(\sigma|\mathcal{I} \right)  p\left( \mathbf{I_{obs}} | a,\mathbf{R_f}, \mathbf{P},\mathbf{Y_0},\mathbf{\sigma} ,\mathcal{I}\right), 
%\end{align}

\begin{align}
	p\left(a,\sigma | \mathbf{I_{obs}},\mathbf{R_f}, \mathbf{P},\mathbf{Y_0}, \mathcal{I} \right) &\propto p\left(a,\sigma| \mathbf{R_f},\mathcal{I}\right) p\left( \mathbf{I_{obs}}|p,\sigma , \mathbf{R_f}, \mathbf{P},\mathbf{Y_0}, \mathcal{I}\right)\notag  \\
	&\propto p\left(a|\sigma, \mathbf{R_f},\mathcal{I}\right)p\left(\sigma|\mathcal{I} \right)  p\left( \mathbf{I_{obs}}|p,\sigma , \mathbf{R_f}, \mathbf{P},\mathbf{Y_0}, \mathcal{I}\right)
\end{align}
{where $\mathcal{I}$ denotes any background information written according to the standard Bayesian setup and $\sigma$ is the error variance.} 
\subsection{Prior distributions}
The rainfall coefficient is the parameter relating the rainfall to the per-capita vector density. There is no existing knowledge of this relation and thus we assume a uniform prior for $p\left( a|\mathbf{R_f},\sigma, \mathcal{I}\right)$. 
{The term $P(\mathbf{R_f}|\sigma,\mathcal{I})$ denotes the prior distribution of rainfall. Since there is no existing knowledge about the seasonal rainfall, a uniform distribution is assumed here as well.} However, the existing knowledge on error variance is given by the conjugate prior of the likelihood - the inverse gamma distribution,
\begin{align}
p\left(\sigma|\mathcal{I} \right)  &= p\left(\sigma|\alpha,\beta, \mathcal{I} \right) \notag \\
&=\dfrac{\beta^{\alpha}}{\Gamma\left(\alpha \right)\left(1/\sigma \right)^{\left(\alpha+1 \right) } e^{\left(-\beta/\sigma \right) }  },
\end{align}
where $\alpha,\beta$ are hyper-parameters. Here $\alpha = s^2,\beta=1$ were used in the simulation process where $s$ is the sample standard deviation of $\mathbf{I_{obs}}$ \cite{haario2006dram}.
\subsection{Likelihood distribution}
By the assumption that errors in observed dengue data $\mathbf{I_{obs}}$ follows Gaussian distribution with variance $\sigma^2$, the likelihood distribution for a season is given below.
\begin{align}\label{likelihood}
	\log p\left( \mathbf{I_{obs}} | a,\mathbf{R_f}, \mathbf{P},\mathbf{Y_0},\mathbf{\sigma} ,\mathcal{I}\right) = \dfrac{n}{2} \log \left(\dfrac{1}{2\pi \sigma^2}\right)- \dfrac{1}{2\sigma^2} 
	\|{\mathbf{I_{obs}} - \mathbf{\hat{I}}}\|^2,   
\end{align}
{where $||.||^2$ denotes the $L2$ norm of the errors.} 
\subsection{Posterior distribution}
The resulting posterior distribution can be summarized as,
\begin{align}
&\log p\left( a,\sigma | \mathbf{I_{obs}},\mathbf{R_f}, \mathbf{P},\mathbf{Y_0}, \mathcal{I}\right) \propto \notag \\
&\log p\left(\sigma|\alpha,\beta, \mathcal{I} \right) + \dfrac{n}{2} \log \left(\dfrac{1}{2\pi \sigma^2}\right)- \dfrac{1}{2\sigma^2} 
\|{\mathbf{I_{obs}} - \mathbf{\hat{I}}}\|^2
\end{align}
%\begin{align}
%	&\log P(\mathbf{z}|\mathbf{I_{obs}},\mathbf{P},\mathbf{R_f},\sigma,\mathcal{I}) = \notag \\
%	& K+ \log P(\mathbf{z}|\mathbf{R_f},\mathcal{I}) + \log P(\mathbf{R_f}|\mathcal{I}) + \dfrac{n}{2} \log \left(\dfrac{1}{2\pi \sigma^2}\right)- \dfrac{1}{2\sigma^2} 
%	\|{ I_{obs(i)} - \hat{I}_i}\|^2   
%\end{align}
We then encapsulated this Bayesian posterior simulation process into the Alg. \eqref{algo1} showcasing the full system of parameter estimation for all seasons, $p$. The optimization problem is a posterior simulation with MCMC while minimizing the error from which a minimum error will always be found regardless of the magnitude of the error. The error between the data and the simulation results ($\sum \epsilon^2$) are taken after solving the ODE system in Eq.\eqref{3D_syst} from which a numerical solution will always be obtained.
\begin{algorithm}
	\caption{Estimating seasonal rainfall coefficients, $a_i,\quad i=1,\cdots,p$} 
	\begin{algorithmic}[1]
		%give the fix parameters that will not change with seasons
		\State {Input $\mathbf{I_{obs}},\mathbf{R_f},\mathbf{P},\mathbf{Y_0}$}
		\State Set $y_{initial}$ =$ \mathbf{Y_0}\left( 1\right) $
		\State Set $a\left( 0\right) =\mathbf{Y_0}(\text{end})$
	%	\State Solve model \eqref{3D_syst} with yinitial,$\mathbf{R_f},\mathbf{P}$
		\For {$i=1:p$} 
		\State Solve ODE in Eq. \eqref{3D_syst} for $\mathbf{\hat{I}}$ with $a\left( 0\right)$ 
		\State Simulate the posterior,  $\log p\left( a_i,\sigma_i| \mathbf{I_{obs}},\mathbf{R_f}, \mathbf{P},\mathbf{Y_0}, \mathcal{I}\right)$
		\State Set E$\left( \log p\left(a_i|\mathbf{I_{obs}},\mathbf{R_f}, \mathbf{P},\mathbf{Y_0},\mathbf{\sigma}, \mathcal{I} \right) \right) =a_i$
			\State Set E$\left( \log p\left(\sigma_i|\mathbf{I_{obs}},\mathbf{R_f}, \mathbf{P},\mathbf{Y_0},\mathbf{\sigma}, \mathcal{I} \right) \right) =\sigma_i$
		\State Set $y_{initial}$ = $\mathbf{\hat{I}}(\text{end})$
		\State Set $a\left(0 \right) =  a_i$
		\EndFor 
		\State Output $\mathbf{a^*}=[a_1,...,a_p],\mathbf{\sigma^*}=[\sigma_1,\sigma_2,\cdots,\sigma_p]$  
	\end{algorithmic} 
	\label{algo1}
\end{algorithm}

\section{Results \& Discussion}
\begin{figure}[ht!]
	\includegraphics[width=\textwidth]{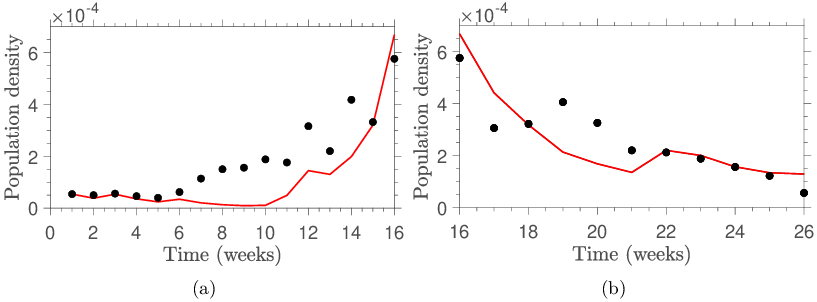}
	\caption{Parameter estimation of the first two seasons. The black dots indicate the reported dengue infected population density and the red line indicates the estimated line. The fitted curve in (a) is obtained taking the actual initial values at the beginning of the time period. The parameter value obtained for this season is 10.7882. The fitted curve in (b) is obtained by taking the initial values to be the final values of the fitted curve in (a). The parameter value obtained for the second season is 2.131.}
	\label{fig_4}
\end{figure}

\begin{figure}[ht!]
	\includegraphics[width=\textwidth]{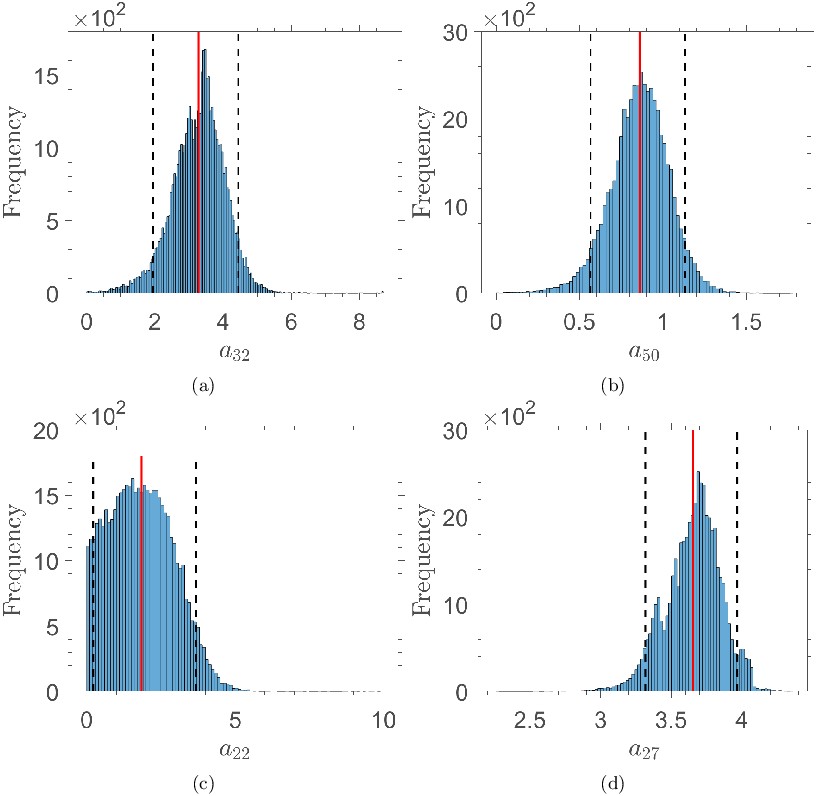}
	\caption{Histograms of the marginal distribution obtained for 4 seasons: (a) 2017 January-April (b) 2021 April-July, (c) 2009 July-October, (d) 2015 October-January. The red vertical line indicate the mean of the marginal distribution. The black dotted lines indicate the 95\% credible intervals calculated for each marginal distribution.}
	\label{fig_5}
\end{figure}

In the simulations of model \eqref{3D_syst}, the initial values for $\textbf{I},\textbf{R},\textbf{V}$ for the very first season were taken respectively as [$I_0,0,V_0$] where $V_0$ is calculated assuming the vector population is at the quasi-equilibrium. The parameters of the model \eqref{3D_syst} denoted by $\mathbf{P}$ are specific to dengue disease and were extracted from literature. The values taken for this study are $\beta_h= 0.75, \mu_h= \frac{1}{75\hspace{0.1cm} \text{years}}, \beta_v=0.375, \gamma_h = \frac{1}{2\hspace{0.1cm} \text{weeks}}$ \cite{derouich2003model} and $\mu_v = \frac{1}{6 \hspace{0.1cm}\text{weeks}}$ \cite{vdciMosquitoBiology}. Starting from the second season, the initial values are derived from the estimated solutions of the preceding season. Fig. \ref{fig_4} shows the solution curves of $\mathbf{\hat{I}}$ against observed data ($\mathbf{I_{obs}}$) for the first two seasons. The solution curve from the model in Eq.\eqref{3D_syst} is obtained for each season by solving the optimization problem in Alg. \ref{algo1} using the MCMC simulations. For each season $50000$ simulations were sufficient to observe the convergence of the samples. With these obtained convergent samples $30\% $ of samples were burnt for accuracy purposes and the marginal probability distributions were obtained for each seasonal parameter. Marginal distributions obtained for four seasons are depicted in Fig. \ref{fig_5}. The best estimate is taken as the mean of the marginal distribution.

\begin{figure}[ht!]
	\centering
	\includegraphics[width=\textwidth]{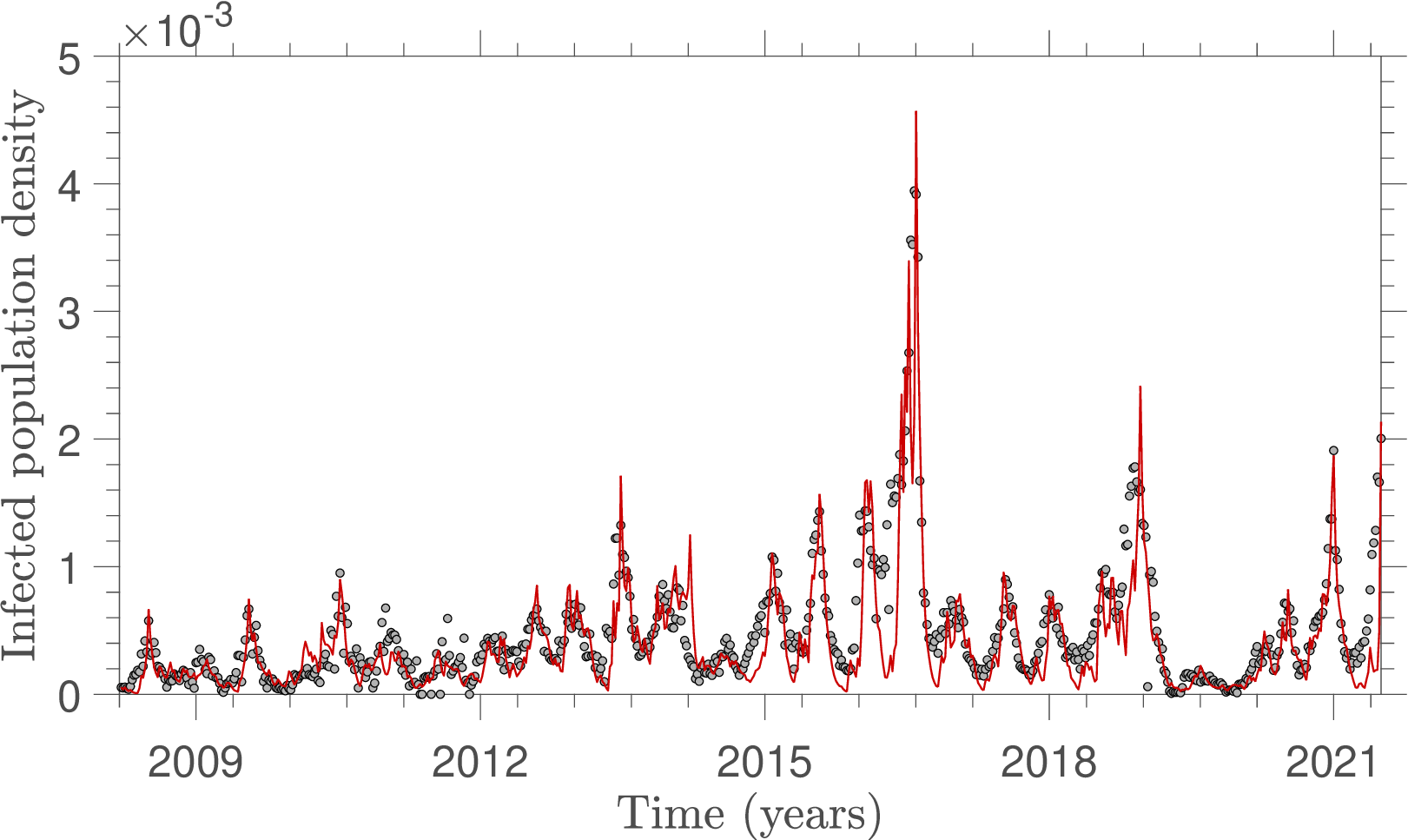}
	\caption{Estimated dengue infected human population density in the CMC area: the red line shows the estimated results (See Table \ref{tab3}). The gray data point show the reported dengue infected human population density.}
	\label{fig_6}
\end{figure}

\begin{figure}[ht!]
	\centering
	\includegraphics[width=\textwidth]{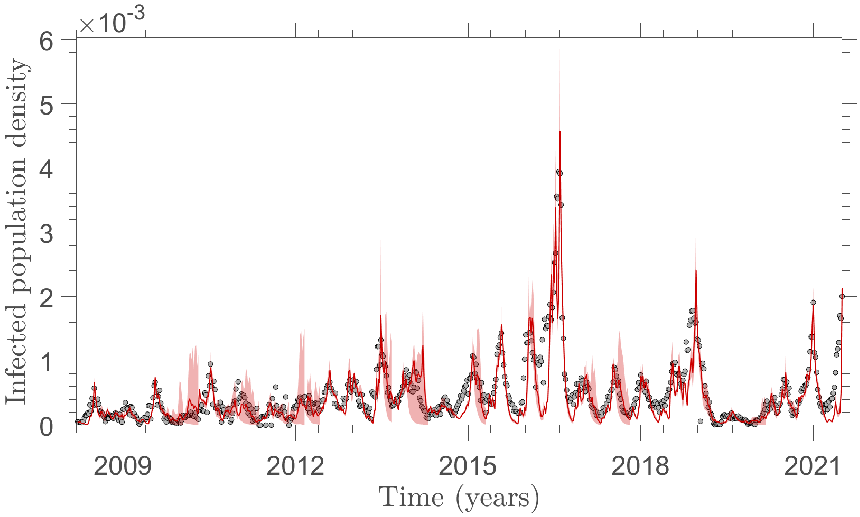}
	\caption{Uncertainty of results: The red line shows the estimated results. The pink color band shows the uncertainly band for each season.}
	\label{fig_7}
\end{figure}

 By using the endpoint of estimated $I$ as the initial value $\left( \hat{I}(0)\right) $ for the subsequent season, the iterative estimations of infected human population density for all 54 seasons were performed. The comparison of estimated dengue cases vs reported dengue cases are depicted in Fig. \ref{fig_6}. The results were able to accurately capture the dengue outbreaks which took place during 2017 and 2019. With the information of initial values, per-capita vector density and the rainfall for a particular season, the algorithm was able to forecast the dengue cases in a consistent periodicity. The uncertainty of these estimates were obtained by calculating the credible intervals at 95\% level from the simulated marginal distributions of rainfall coefficient (see Fig. \ref{fig_5}, \ref{fig_7}).

However, the estimated results shown in Fig. \ref{fig_7} shows large uncertainties in specific seasons which needs further investigation. The marginal distributions of these seasons show the possible existence of mix Gaussian distribution instead of a Gaussian distribution suggesting the impact of some external factors that may overpower the impact of rainfall (Fig. \ref{fig_5}-(c)). Especially during the festive seasons, disease transmission elevates due to the increasing human mobility and population gatherings. In addition, unexpected above-average rainfall incidences also may have an impact in the differences shown in the marginal distributions.  

\begin{figure}[htp]
	\centering
	\includegraphics[width=\textwidth]{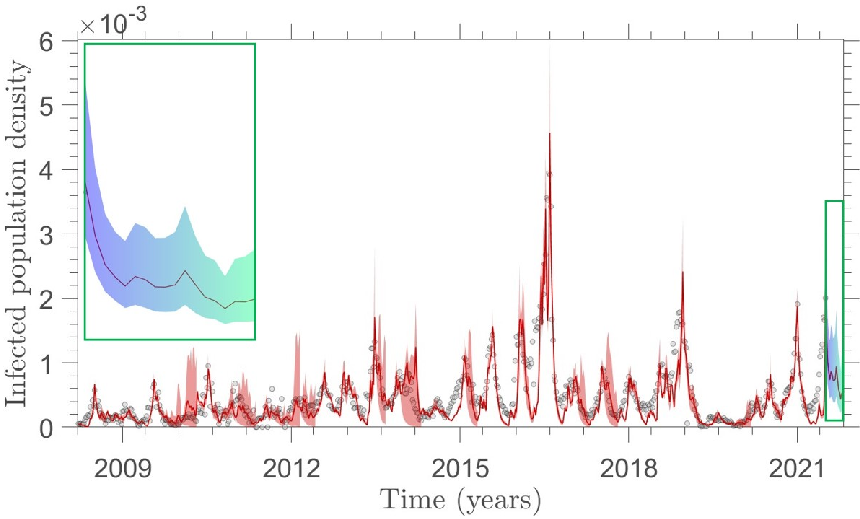}
	\caption{Forecast of dengue emergence using the average seasonal rainfall coefficient: dengue emergence forecast for one upcoming year is given in the green box. The red line shows the estimated forecast and the shaded region gives the uncertainty band for the recorded initial value. }
	\label{fig_8}
\end{figure}

In this study, in which the parameter estimation is done categorically for four seasons within a year, a common parameter value for each season is identified by taking the average marginal distribution. By utilizing these parameter values, one can forecast upcoming dengue emergence only using the rainfall data (see Fig. \ref{fig_8}). {Thus, in this study an out-of-sample forecast was employed where dengue emergence is forecasted for a period of 10 upcoming weeks using rainfall data.} The uncertainty of this forecast is indicated in the colored region.  

\section{Conclusions}

Sri Lanka experiences a severe health threat from the dengue mosquitoes for many years and the only way for survival from this threat is to try and mitigate the spread of the disease from infected population and infected vectors. Therefore the capacity to forecast the dengue cases can help the relevant authorities to handle the situation in advance with their disease curbing pursuits. However, forecasting of the infected population is only addressing the top layer of the problem. The actuality of the problem is coming from infected vectors. Thus trying to forecast the infected vector population is superior helping the authorities to avoid spreading the disease and more importantly can avoid generating infected vectors through infected population. In this study, we estimate the per-capita vector density using rainfall information. Sri Lanka is a tropical country that experiences seasonal rainfall throughout the year. We believe the seasonality of rainfall affect the breeding of vectors and thus increasing the probability of generating infected vector population. This dynamic nature of rainfall is used in our procedure of estimating dengue infected vectors and population. 

In particular we conduct our study in the CMC area, where the intensity of population is very high compared to other regions in the country with the highest number of reported dengue cases. The dynamic nature of dengue incidences were captured via feeding the rainfall information into the mathematical model in Eq. \eqref{3D_syst}. In that, we estimated the per-capita vector density as a dynamic parameter depending on identified seasons. The per-capita vector density is a dynamic parameter that depends on the rainfall information. We estimate a seasonal rainfall coefficient that reflects the dependency of the rainfall to the per-capita vector density (see Eq. \eqref{z_with_Rf}). This estimation procedure is performed as an iterative algorithm (see Alg. \ref{algo1}) by incorporating a Bayesian parameter estimation paradigm \citep{haario2006dram}. The parameter estimation is performed via simulating marginal probability distributions of rainfall coefficient for each season (54 seasons were identified in the study period) with MCMC (see Table. \ref{tab3}). 

The dengue cases estimated from our algorithm are closely following the recorded cases. In particular, the estimations were able to capture the dengue outbreaks that took place during 2017 and 2019. It indicates that the factors affecting to create an outbreak was embedded in the rainfall information, initial value of infected population value for the season and more importantly the per-capita vector density. It shows that ability to predict the dynamic per-capita vector density is very crucial in making the estimations successful. The Bayesian paradigm used in our algorithm enabled to quantify the uncertainty of the estimates (see Fig. \ref{fig_7}). The uncertainty of estimates revealed further non-obvious information. In fact, the uncertainty of 2017 and 2019 outbreaks were very low which validates the accuracy of the estimates of an outbreak. The results of this study also confirms that the rainfall is the dominant factor in deciding the dengue emergence in a given season. Further, the uncertainty in some seasons were unrealistically very high. Since this study is conducted under the assumption that the rainfall governs the per-capita vector density and hence the dengue disease emergence, these observed high uncertainties (Fig \ref{fig_7}) suggest the possibility of other external factors that may overpower the effect of rainfall on disease emergence. Hence further investigation of the uncertainties is crucial.

%These uncertainties are recorded in national festive seasons and in periods of unexpected climate changes.

Upon careful observation it can be seen that the inferred high uncertainties are recorded in many of the seasons following the Christmas/New Year season in Sri Lanka. Colombo Municipal Council (CMC) area being the commercial city of the country, the Christmas and New year seasons makes it the busiest and the most crowded city during these times. Due to the high population density with their inbound and outbound mobility and the festivities, the city's environmental pollution elevates as well. Thus during this season, the impact of rainfall is therefore overpowered which verify the marginal with high uncertainty. The rest of the seasons that have large uncertainties are reported following the seasons having above-average rainfall. Hence it can be concluded that the external factors that may overpower the impact of rainfall may produce uncertainties in the forecast. Therefore, one can extend the method introduced in this study to forecast the dengue emergence while considering these external factors as well.

%This is caused partially by the error accumulated in the estimated initial value of dengue cases for each season. Moreover the high unexpected uncertainties are caused by any hidden uncertainty that would propagate throughout seasons in the rainfall information which would need further investigation in the paradigms of uncertainty propagation. 

Overall, we present an iterative algorithm that estimates the seasonal dengue cases using a probabilistic and dynamic modeling approach. We showcased the superiority in capturing outbreaks of dengue and validate it using actual data and uncertainty quantification. We also identified the challenges in dynamic parameter estimation in terms of uncertainty blowups which would need further future investigations. 
\clearpage
\section{Appendix}
	\begin{figure}[ht]
	\centering
	\begin{subfigure}{0.48\textwidth}
		\includegraphics[width=\linewidth]{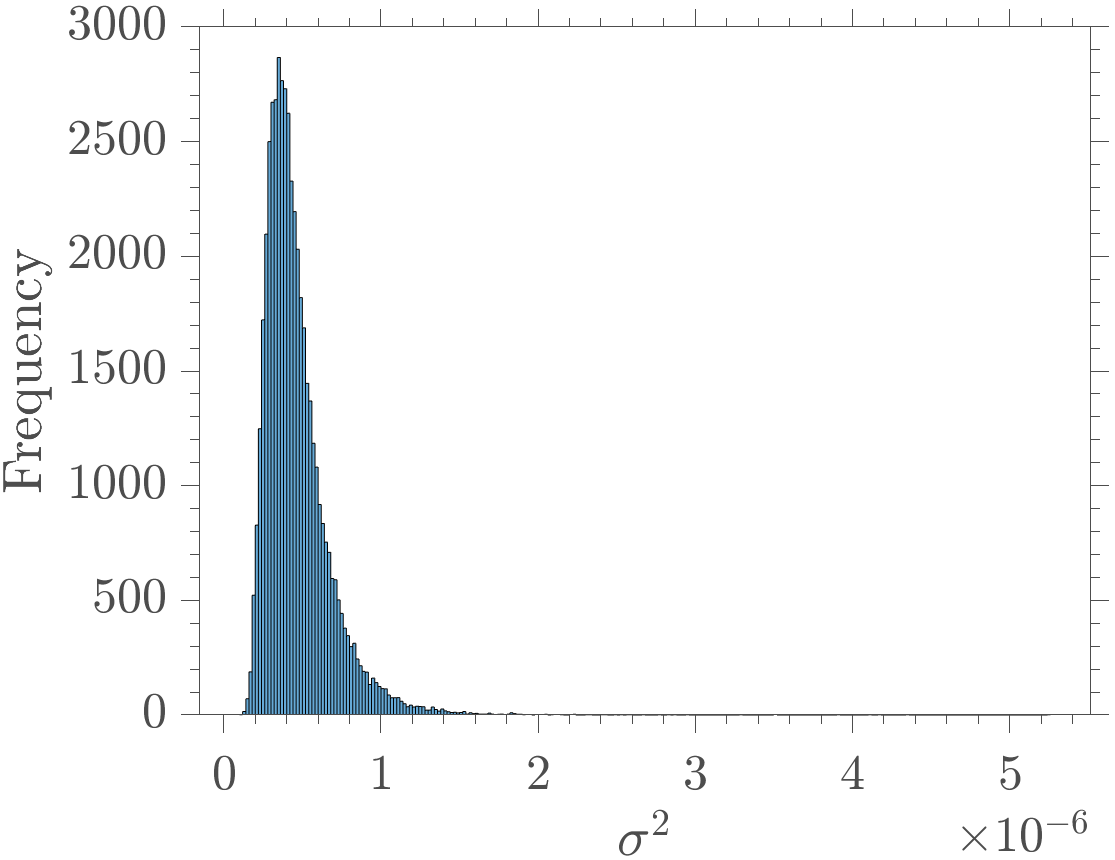}
		\caption{}
	\end{subfigure}
	\begin{subfigure}{0.48\textwidth}
		\includegraphics[width=\linewidth]{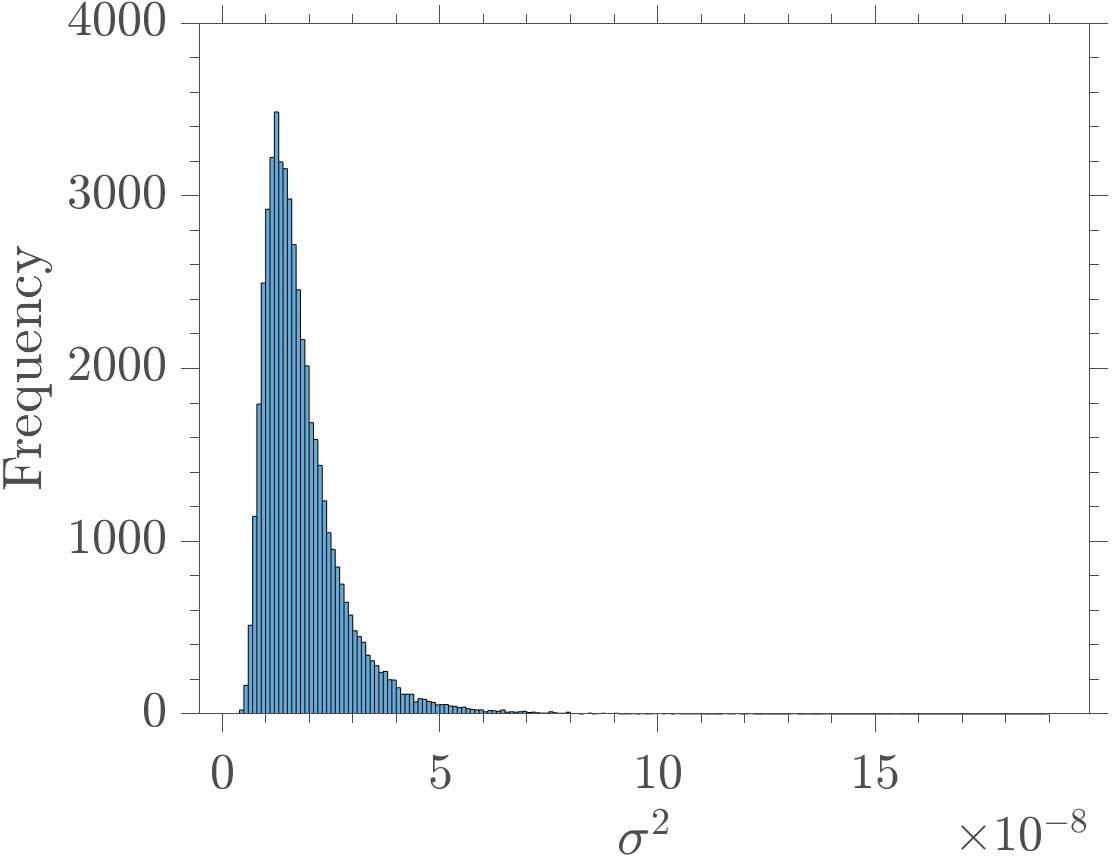}
		\caption{}
	\end{subfigure}
	\begin{subfigure}{0.48\textwidth}
		\includegraphics[width=\linewidth]{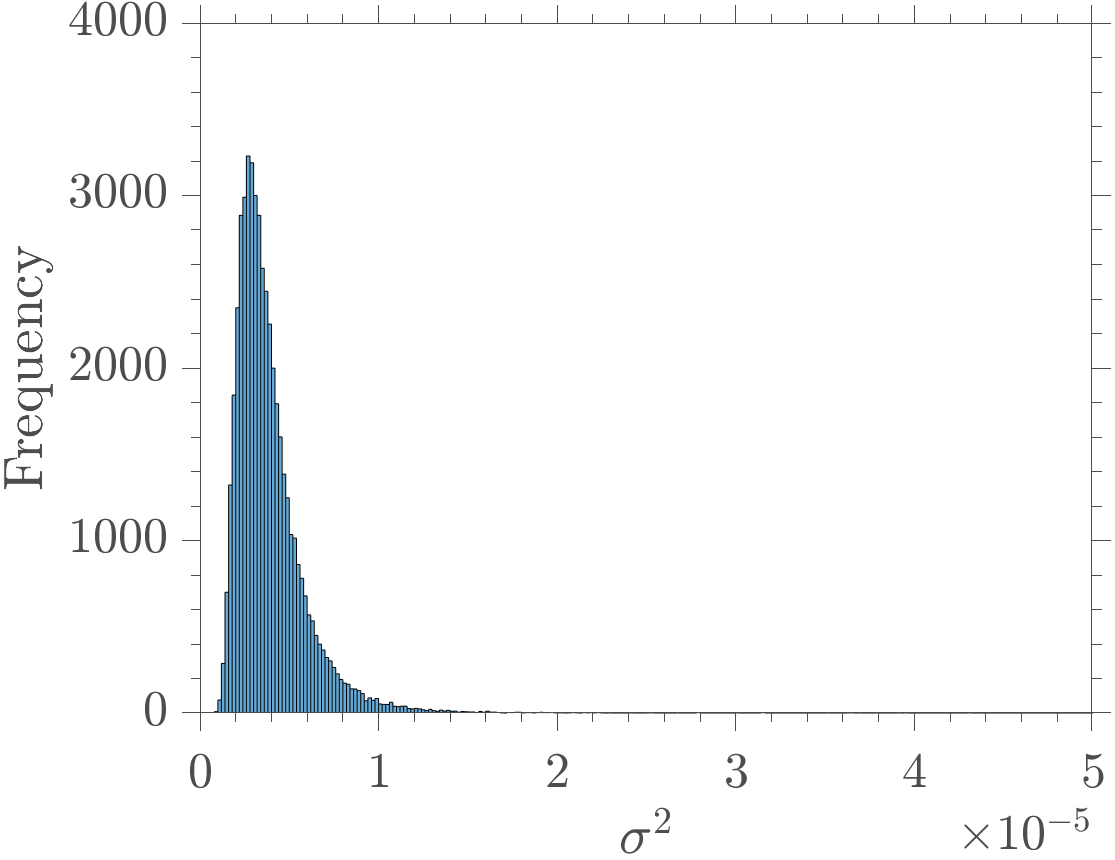}
		\caption{}
	\end{subfigure}
	\begin{subfigure}{0.48\textwidth}
		\includegraphics[width=\linewidth]{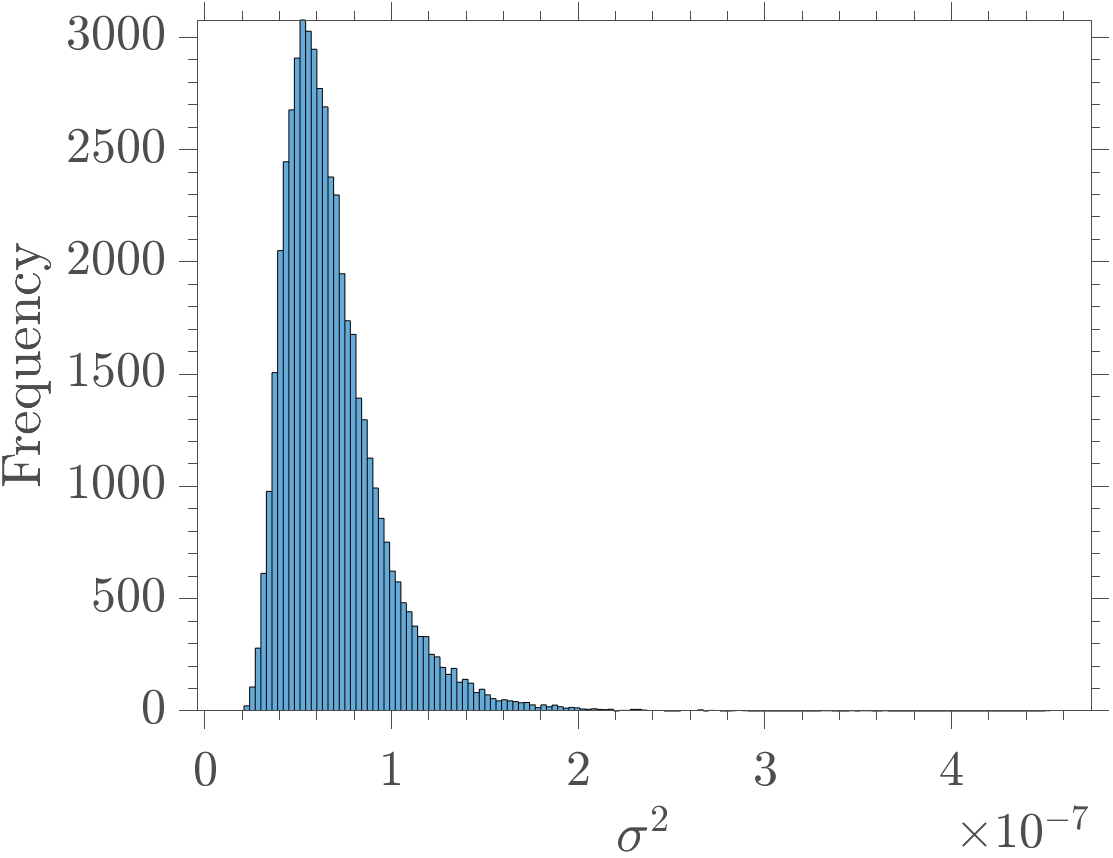}
		\caption{}\end{subfigure}
	\caption{Marginal distribution of variance of the model for four seasons: (a) 2017 January-April (b) 2021 April-July, (c) 2009 July-October, (d) 2015 October-January.}
	
	\label{fig:panel}
\end{figure}
\clearpage
 
	\begin{table}[ht!]
	\centering
	\caption{Estimated parameter values of each season with the confidence intervals}
	\label{tab3}
	\begin{tabular}{@{}lllll@{}}
		\hline
		Season & Time Span & Parameter Value & 95\% confidence interval & SSE \\
		\hline
		1 & 1--16 & 10.78 &[ 9.19 , 12.14] & 5.0E-07 \\
		2 & 16--26 & 2.13 &[ 0.90 , 3.20 ]& 8.2E-07 \\
		3 & 26--44 & 3.51 &[ 2.81 , 4.09 ]& 3.4E-07 \\
		4 & 44--57 & 1.79 & [1.36 , 2.19 ]& 2.5E-07 \\
		5 & 57--71 & 7.17 &[ 6.42 , 7.67 ]& 7.5E-07 \\
		6 & 71--81 & 1.13 &[ 0.82 , 1.40 ]& 1.3E-06 \\
		7 & 81--96 & 1.85 &[ 0.09 , 4.22 ]& 8.6E-07 \\
		8 & 96--110 & 2.34 &[ 0.10 , 5.10 ]& 5.6E-07 \\
		9 & 110--121 & 4.57 &[ 3.70 , 5.21 ]& 1.2E-06 \\
		10 & 121--131 & 1.04 &[ 0.39 , 1.64 ]& 2.2E-06 \\
		11 & 131--146 & 4.70 &[ 2.79 , 6.04 ]& 1.5E-06 \\
		12 & 146--165 & 2.10 &[ 0.12 , 4.18 ]& 1.2E-06 \\
		13 & 165--180 & 3.92 &[ 2.74 , 4.77 ]& 3.8E-07 \\
		14 & 180--192 & 3.42 &[ 0.68 , 5.58 ]& 5.3E-07 \\
		15 & 192--206 & 2.51 &[ 0.12 , 5.39 ]& 5.7E-07 \\
		16 & 206--220 & 2.37 &[ 0.19 , 4.47 ]& 4.1E-07 \\
		17 & 220--229 & 4.72 &[ 4.03 , 5.27 ]& 7.3E-07 \\
		18 & 229--240 & 1.59 &[ 1.40 , 1.77 ]& 1.1E-06 \\
		19 & 240--250 & 6.15 &[ 4.98 , 7.09 ]& 1.3E-06 \\
		20 & 250--266 & 2.18 &[ 1.91 , 2.40 ]& 1.7E-06 \\
		21 & 266--275 & 23.86 & [16.47 , 28.51] & 3.2E-06 \\
		22 & 275--285 & 1.84 &[ 0.11 , 3.99 ]& 6.2E-06 \\
		23 & 285--298 & 4.36 &[ 3.87 , 4.83 ]& 3.3E-06 \\
		24 & 298--318 & 1.91 &[ 0.09 , 4.16 ]& 2.1E-06 \\
		25 & 318--333 & 2.21 &[ 1.97 , 2.48 ]& 4.5E-07 \\
		26 & 333--342 & 3.47 &[ 2.47 , 4.43 ]& 1.2E-07 \\
		27 & 342--358 & 3.65 &[ 3.24 , 4.02 ]& 1.0E-06 \\
		28 & 358--370 & 1.64 &[ 0.12 , 3.43 ]& 3.7E-06 \\
		29 & 370--384 & 8.37 &[ 7.85 , 8.94 ]& 4.9E-06 \\
		30 & 384--399 & 0.71 &[ 0.40 , 0.98 ]& 1.2E-05 \\
		31 & 399--409 & 19.49 & [12.82 , 23.49] & 3.5E-06 \\
		32 & 409--422 & 3.27 &[ 1.58 , 4.67 ]& 1.6E-05 \\
		33 & 422--437 & 9.79 &[ 8.56 , 10.66] & 2.1E-05 \\
		34 & 437--449 & 0.41 &[ 0.07 , 0.77 ]& 8.7E-05 \\
				\hline
		\end{tabular}
	\end{table}
	
		\begin{table}[ht!]
		\centering
		%\caption{Estimated parameter values of each season with the confidence intervals}
		%\label{tab3}
		\begin{tabular}{@{}lllll@{}}
			\hline
			Season & Time Span & Parameter Value & 95\% confidence interval & SSE \\
			\hline
		
		35 & 449--461 & 3.54 &[ 2.90 , 4.13 ]& 4.8E-06 \\
		36 & 461--474 & 1.34 &[ 0.07 , 2.77 ]& 6.1E-06 \\
		37&	474--486&	7.62&	[6.59 ,	8.37]&	1.5E-06\\
		38&	486--500&	1.55&	[0.09 ,	3.46]&	3.1E-06\\
		39&	500--512&	6.04&	[4.78 ,	7.00]&	1.5E-06\\
		40&	512--527&	2.30&	[1.75 ,	2.74]&	2.5E-06\\
		41&	527--539&	9.62&	[8.07 ,	10.95]&	1.1E-06\\
		42&	539--549&	7.01&	[5.29 ,	8.55]&	1.2E-06\\
		43&	549--560&	4.10&[	3.13 ,	4.82]&	4.2E-06\\
		44&	560--576	&0.62&[	0.33 ,	0.89]&	1.6E-05\\
		45&	576--592&	4.17&	[2.27 ,	5.53]&	2.3E-06\\
		46&	592--607&	2.02&[	1.45 ,	2.51]&	1.1E-07\\
		47&	607--623	&3.36&[	0.29 ,	5.59]&	7.0E-08\\
		48&	623--633&	7.23&	[5.62 ,	8.63]&	4.2E-07\\
		49&	633--641&	7.14&[	5.99 ,	8.51]&	3.0E-07\\
		50&	641--651&	0.86&[	0.49 ,	1.19]&	1.3E-06\\
		51&	651--666&	5.89&[	5.60 ,	6.06]&	5.0E-06\\
		52&	666--679&	0.52&[	0.34 ,	0.69]&	1.1E-05\\
		53&	679--692&	5.37&[	3.30 ,	6.60]&	4.2E-06\\
		54&	692--700&	1.58&[	1.21 ,	1.88]&	6.6E-06\\
		
		\hline
	\end{tabular}
\end{table}

\clearpage

%%% The Appendices part is started with the command \appendix;
%%% appendix sections are then done as normal sections
%\appendix
%
%\section{Sample Appendix Section}
%Lorem ipsum dolor sit amet, consectetur adipiscing elit, sed do eiusmod tempor section \ref{sec:sample1} incididunt ut labore et dolore magna aliqua. Ut enim ad minim veniam, quis nostrud exercitation ullamco laboris nisi ut aliquip ex ea commodo consequat. Duis aute irure dolor in reprehenderit in voluptate velit esse cillum dolore eu fugiat nulla pariatur. Excepteur sint occaecat cupidatat non proident, sunt in culpa qui officia deserunt mollit anim id est laborum.

 %\bibliographystyle{elsarticle-num} 
 %\bibliographystyle{unsrt}
 %\bibliography{bibliography}





% \end{thebibliography}
\end{document}